\documentclass[twocolumn,aps]{revtex4}
\usepackage[latin9]{inputenc}
\usepackage{amsmath}
\usepackage{amssymb}
\usepackage{graphicx}

\makeatletter
\@ifundefined{textcolor}{}
{%
 \definecolor{BLACK}{gray}{0}
 \definecolor{WHITE}{gray}{1}
 \definecolor{RED}{rgb}{1,0,0}
 \definecolor{GREEN}{rgb}{0,1,0}
 \definecolor{BLUE}{rgb}{0,0,1}
 \definecolor{CYAN}{cmyk}{1,0,0,0}
 \definecolor{MAGENTA}{cmyk}{0,1,0,0}
 \definecolor{YELLOW}{cmyk}{0,0,1,0}
}

\newcommand{\ket}[1]{\ensuremath{|{#1\rangle}}} 
\newcommand{\bra}[1]{\ensuremath{{\langle #1}|}}

\usepackage{amsfonts}\usepackage{color}\usepackage{epsfig}\usepackage{bm}\usepackage{rotating}\usepackage{mathrsfs}

\makeatother

\begin{document}

\title{Is the Preferred Basis selected by the environment?}

\author{Tian Wang}
\email{tianwang@ucalgary.com}

\affiliation{Institute for Quantum Science and Technology, University of Calgary,
Alberta T$2$N $1$N$4$, Canada}
 
 \author{David Hobill}
 \affiliation{Institute for Quantum Science and Technology, University of Calgary,
Alberta T$2$N $1$N$4$, Canada}
 
\begin{abstract}
We show that in a quantum measurement, the preferred basis is determined
by the interaction between the apparatus and the
quantum system, instead of by the environment. This interaction entangles
three degrees of freedom, one system degree of freedom we are interested
in and preserved by the interaction, one system degree of freedom
that carries the change due to the interaction, and the apparatus
degree of freedom which is always ignored. Considering all three degrees
of freedom the composite state only has one decomposition, and this
guarantees that the apparatus would end up in the expected preferred
basis of our daily experiences. We also point out some problems with
the environment-induced super-selection (Einselection) solution to the preferred basis problem,
and clarifies a common misunderstanding of environmental decoherence
and the preferred basis problem. 
\end{abstract}
\maketitle

\section{Introduction}

The preferred basis problem (PBP) is one of the most fundamental problems
in the foundations of quantum theory\cite{schlosshauer_decoherence_2005,zurek_decoherence_2003,zurek_environment-induced_1982,zurek_preferred_1993,zurek_pointer_1981}.
It has been one of the key problems associated with the interpretation of quantum
mechanics, in particular the many world interpretation\cite{barvinsky_preferred_1995,barvinsky_preferred_1990}.
The suggestion that the preferred basis problem is solved by the
``environment induced superselection (Einselection)'', was developed
by Zurek \cite{zurek_pointer_1981,zurek_environment-induced_1982}.
This theory maintains that the preferred basis is selected by the environment.
However, the introduction of the environment is not really necessary to solve
the PBP. In addition it can lead to several issues. In this paper we
suggest a different solution to the PBP, that the preferred basis is determined
by the configuration of the measurement apparatus, and the interaction
between the apparatus and the environment only determines whether
the correlation between the system and apparatus is preserved.

As we know from Newton's third law of motion, when one body exerts
a force on a second body, the second body simultaneously exerts a
force equal in magnitude and opposite in direction to that of the
first body. (This gives us momentum conservation, which can equivalantly be
formulated in Lagrangian or Hamiltonian approach.) 
However, when talking about the
quantum measurement process, beginning with von Neumann's contributions,
it is generally supposed that the interaction between the quantum system
and classical apparatus only changes the state of the apparatus, while
leaving the state of the system untouched\cite{zurek_pointer_1981,schlosshauer_decoherence_2005}. 

\begin{equation}
\ket{s_{n}}\ket{a_{0}}\stackrel{H_{sa}}{\longrightarrow}\ket{s_{n}}\ket{a_{n}}
\end{equation}

Here $|s_{n}\rangle$ and $|a_{n}\rangle$ represent the quantum system
and the classical apparatus. This comes from the third postulate of
quantum mechanics, that repetitive measurements yield the same results\cite{vonNeumann}.
However, this directly conflicts with the Newton's third law, and
the fundamental momentum conservation law that can be derived from that. To
solve such a contradiction between classical and quantum mechanics there
are two possible solutions. The first is to regard the effect on the
quantum system is being too small and thus the state of the system
can be treated as unchanged. However, this approximation is problematic.
During the interaction between two objects, the change of the motion
of the object with the smaller mass will be larger than the the more massive
one, as result of the conservation of momentum. In the quantum measurement
scenario, when an interaction is strong enough to alter the classical
apparatus, which is assumed to be far larger than the quantum system,
so as to present a different reading, it is difficult to believe that
the state of the quantum system will be unchanged.

The second solution, as will be discussed in detail in the next section,
is that the quantum system is described by two uncoupled degrees of
freedom. The interaction only changes one degree of freedom, while
the other degree of freedom --the one we want to measure, is preserved. Actually
it turns out that the ordinary examples of quantum measurement such
as Stern-Gerlach (S-G) experiment fall into this category. However,
the degree of freedom describing the macroscopic apparatus is usually
ignored. In our approach this hidden degree of freedom solves the
PBP.

This paper is organized in the following method: Section II introduces
the hidden degree of freedom. Section III briefly introduces the PBP,
the Einselection solution, and our solution that that the preferred
basis is determined by the configuration of the measurement apparatus.
Section IV compares the two solutions, pointing out the advantages
of our solution, and clarifies a common misunderstanding of decoherence
and the role of environment. Finally Section V analyzes the role of the environment
in PBP and the relationship with decoherence.

\section{The hidden degree of freedom}

We begin with the archetype of quantum measurement--the S-G experiment.
A spin one-half particle interacts with a macroscopic set of magnets.
The particle is described by two uncoupled degrees of the freedom,
the spin degree of freedom $\ket{s_{n}}$, which is preserved by the
interaction with the magnet, and the momentum (spatial) dependent
degree of freedom $\ket{p_{n}}$ , which changes in accordance with
$\ket{s_{n}}$, in that when the particles pass through the magnet, spin up particles
will be deflected upwards, and spin down particles deflected downwards,
both by specific amounts. (It should be noted that $\ket{p_{n}}$ are not the exact momentum eigenstates.) However, according to Newton's third law
the interaction between the particle and magnet should not only change
the state of the particle, but also the state of the magnet, which
is always ignored in the description of this experiment.

Usually, the position of the particle is considered as the apparatus. However, according to the measurement scheme, the apparatus
should be some macroscopic object, the change of which can be read
by us easily, namely, we do not need to perform an extra quantum measurement
to determine the state of the apparatus. From the point of view of
the foundations of quantum mechanics, the apparatus is usually
illustrated as a panel with a pointer, which can be read by a human
observer. In the S-G experiment, it is the macroscopic magnet that
should be regarded as the apparatus. The interaction between the particle
and the magnet changes the direction of the velocity of the particle
according to which spin it carries. The same interaction will also
have an effect on the magnet: the magnet is subject to a torque whose
direction and magnitude depend on the particle's spin as it interacts
with the magnet.
If the system and apparatus are well separated from
the environment, and the magnet is delicate enough, the magnet will
rotate in different directions with different spin inputs. The spatial
degree of freedom of the particle is actually the one that carries
the change due to the interaction, while the spin state is the one
we are interested in and is preserved in the measurement. The composite
system of the particle and the magnet before the measurement will
be 
\begin{equation}
\frac{1}{\sqrt{2}}(\ket{s_{+}}+\ket{s_{-}})\ket{p_{0}}\ket{a_{0}}
\end{equation}
Here $s$ describes different spin, $p$ and $a$ describes the spatial
state of the particle and the state of the magnet respectively. It
will evolve according to 
\begin{equation}
\frac{1}{\sqrt{2}}(\ket{s_{+}}+\ket{s_{-}})\ket{p_{0}}\ket{a_{0}}\stackrel{H_{sa}}{\longrightarrow}\frac{1}{\sqrt{2}}(\ket{s_{+}}\ket{p_{+}}\ket{a_{+}}+\ket{s_{-}}\ket{p_{-}}\ket{a_{-}})
\end{equation}
This is similar to the case in
quantum optics, when the Polarizing Beam Splitter (PBS) is used to
entangle the polarization and the trajectory of a photon. As we know,
a Wollaston prism type PBS will deflect the horizontally polarized
photon toward the left and vertically polarized photon toward the
right, as shown in Fig 1. From momentum conservation we can see that
the PBS will undergo a momentum transfer depending on the different
polarization of the photons deflected. In this case, the polarization
and position state are the states of the two degrees of freedom of
the system, which are preserved and changed respectively. Thus for
an apparatus consisting of a PBS, its momentum which will change differently
according to the polarization of the photon interacting with it.(It
should be noted that a successful PBS that serves the goals for quantum
optics should be large enough so that $\langle a_{H}|a_{V}\rangle\approxeq1$.)

One may argue that the perturbation of the apparatus (whether a magnet
or a PBS) is so small that it is less than their self uncertainty,
and the states are basically the same as before the measurement. This
involves a deep issue of how to interpret the quantum state and uncertainty
of the apparatus, a macroscopic object, which yet to be clarified
(the standard Copenhagen interpretation simply states that classical
objects are not described by QT). In principle, $\ket{a_{H}}$ and
$\ket{a_{V}}$ should
not be exactly the same before and after the interaction, although
the overlap is nearly, but not exactly, 1. This tiny difference has
profound implications and provides us the solution of the preferred
basis problem. 

\begin{figure}

\includegraphics[width=1\columnwidth]{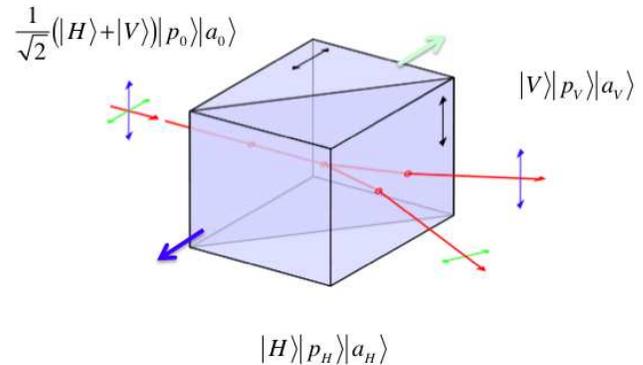}
\caption{\label{fig: demo-state}
 A photon with polarization $\ket{H}$ or $\ket{V}$ will
be deflected by the PBS into left and right directions respectively,
with a spatial mode $\ket{p_{H}}$ and $\ket{p_{V}}$. The recoil
of the PBS in different directions due to different momentum changes
will be described by $\ket{p_{H}}$, $\ket{p_{V}}$ correspondingly.
Consider a photon whose polarization is in a superposition state.
The composite state of the photon and PBS will be $\frac{1}{\sqrt{2}}(\ket{H}+\ket{V})\ket{p_{0}}\ket{a_{0}}$,
with the initial spatial state of the photon $\ket{p_{0}}$ and the
initial state of the PBS $\ket{a_{0}}$. After passing through the
PBS, the spatial degree of freedom of the photon and the recoil of
PBS will be the entangled state $\frac{1}{\sqrt{2}}(\ket{H}\ket{p_{H}}\ket{a_{H}}+\ket{V}\ket{p_{V}}\ket{a_{V}})$
, although the overlap between $\ket{a_{H}}$ and $\ket{a_{V}}$ is
almost 1.}
\label{PBS-fig}

\end{figure}

\section{preferred basis problem}

The ideal measurement scheme was first introduced by von Neumann in
his masterpiece \emph{Mathematische Grundlagen der Quantenmechanik}\cite{vonNeumann}.
A typical microscopic system $\mathcal{S}$, described by basis vectors
$\{\ket{s_{n}}\}$ in a Hilbert space $\mathcal{H}_{\mathcal{S}}$,
interacts with a measuring apparatus $\mathcal{A}$, represented by
basis vectors $\{\ket{a_{n}}\}$ spanning a Hilbert space $\mathcal{H}_{\mathcal{A}}$,
where the $\ket{a_{n}}$ are assumed to correspond to \emph{macroscopically}
distinguishable ``pointer'' positions that in turn correspond to the outcome
of a measurement if $\mathcal{S}$ is in the state $\ket{s_{n}}$.
Now consider the micro-system in an arbitrary state $\{\ket{s_{n}}\}$
interacting with the apparatus in the initial state $\ket{a_{0}}$.
After the measurement interaction, the final state of the apparatus
will become $\ket{a_{n}}$, while leaving the state of the system unchanged.
\begin{equation}
\ket{s_{n}}\ket{a_{0}}\stackrel{H_{sa}}{\longrightarrow}\ket{s_{n}}\ket{a_{n}}
\end{equation}
Here the $\ket{a_{n}}$ are assumed to correspond to macroscopically
distinguishable ``pointer'' positions, $\langle a_{m}|a_{n}\rangle=\delta_{m}{}_{n}$.
Since $\{\ket{s_{n}}\}$ spans a complete Hilbert space, any arbitrary
state $\ket{\psi}$ can be expressed as $\sum_{n}c_{n}\ket{s_{n}}$
where $c_{n}=\langle s_{n}|\psi\rangle$. This can be understood as
the system's state being in a superposition of different $\ket{s_{n}}$,
which is possible for microscopic system. If such a system interacts
with the measuring apparatus, according to the linearity of the Schr\"{o}dinger
equation the total system $\mathcal{SA}$ will evolve as

\begin{equation}
\bigg(\sum_{n}c_{n}\ket{s_{n}}\bigg)\ket{a_{0}}\stackrel{H_{sa}}{\longrightarrow}\sum_{n}c_{n}\ket{s_{n}}\ket{a_{n}}.\label{eq:measurement1}
\end{equation}

The final state of the total system suggests that the system-apparatus
is now in a superposition state. However, our experience tell us that
this macroscopic superposition has never been observed. Either there
should be some physical process that destroys the superposition or
a proper interpretation concerning this superposition is required.
This is the essence of the quantum measurement problem, which should
be divided into two subproblems\cite{schlosshauer_decoherence_2005},
namely the \emph{definite outcome problem} and \emph{the preferred
basis problem}.

The definite outcome problem mainly deals with why we get one definite
single outcome after a measurement instead of observing the total
system in superposition state. Associated with this problem are the
questions: why is there a non-unitary measurement process? how does one interpret
this indeterministic process? how does one understand the probability
associated with the outcomes? Even after some 90 years of debate,
there is still no consensus on the answers to these questions. It
should be noted that the definite outcome problem has not been solved
by the decoherence program, which mainly deals with how the system
couples to the environment and how the environment states evolve into
a complete set of orthogonal states. It explains how the states of
the system lose coherence but not how we get one definite state out
of all possible states.

While the definite outcome problem is usually regarded as the measurement
problem that the majority of the interpretations of quantum mechanics
is aimed at explaining, the preferred basis problem is one that receives
relatively less attention.

The preferred basis problem demonstrates that after the system and
apparatus interact they become entangled 
\begin{equation}
\ket{\Psi}=\sum_{n}c_{n}\ket{s_{n}}\ket{a_{n}},
\end{equation}
When \emph{$c_{n}$ are not distinct}, following from the biorthogonal
decomposition theorem we can in general rewrite the state in terms
of different state vectors, 
\begin{equation}
\ket{\Psi}=\sum_{n}c_{n}^{\prime}\ket{s_{n}^{\prime}}\ket{a_{n}^{\prime}},
\end{equation}
such that the same post measurement state seems to correspond to two
different measurements, that is, of the observables $\widehat{A}=\sum_{n}\lambda_{n}\ket{s_{n}}\bra{s_{n}}$
and $\widehat{B}=\sum_{n}\lambda_{n}^{\prime}\ket{s_{n}^{\prime}}\bra{s_{n}^{\prime}}$
of the system, respectively, although in general $\widehat{A}$ and
$\widehat{B}$ do not commute.

It thus seems that from quantum mechanics we cannot tell which observable(s)
of the system is (are) being recorded, via the formation of quantum
correlations, by the apparatus. This can be stated in a general theorem
\cite{zurek_pointer_1981,zurek_preferred_1993,zurek_environment-induced_1982}:
When quantum mechanics is applied to an isolated composite object
consisting of a system $\mathcal{S}$ and an apparatus $\mathcal{A}$,
it cannot determine which observable of the system has been measured---in
obvious contrast to our experience of the workings of measuring devices
that seem to be ``designed'' to measure certain quantities.

The PBP has been a profound challenge to the interpretation of quantum
mechanics. For example, in von Neumann's ``collapse interpretation'',
the question that remains is why should the composite system collapse
into $\ket{s_{n}}\ket{a_{n}}$ and not $\ket{s_{n}^{\prime}}\ket{a_{n}^{\prime}}$
? Alternatively, in the ``many world interpretation'', one must
ask why we observe a universe containing the composite system in $\ket{s_{n}}\ket{a_{n}}$
instead of a universe containing $\ket{s_{n}^{\prime}}\ket{a_{n}^{\prime}}$
?

It is often believed that the preferred basis problem can be solved
by the Environment induced superselection theory (Einselection) by
Zurek \cite{schlosshauer_decoherence_2005,zurek_pointer_1981,zurek_environment-induced_1982,zurek_preferred_1993}.
Zurek introduces the state of the environment $\mathcal{E}$ which
interacts with the composite system $\mathcal{SA}$. The system-apparatus-environment
state then evolves in the following manner: 
\begin{equation}
\ket{\Phi}=\bigg(\sum_{n}c_{n}\ket{s_{n}}\ket{a_{n}}\bigg)\ket{e_{0}}\stackrel{H_{ae}}{\longrightarrow}\sum c_{n}\ket{s_{n}}\ket{a_{n}}\ket{e_{n}}
\end{equation}
where $\langle a_{m}|a_{n}\rangle=\delta_{mn}$. According to the
tridecompositional uniqueness theorem\cite{elby_triorthogonal_1994},
even if the $c_{n}$ are not distinct, the decomposition of the total
system $\ket{\Phi}$ is unique, as long as $\{\ket{s_{n}}\}$, $\{\ket{a_{n}}\}$
are linearly-independent bases and $\{\ket{e_{n}}\}$ is \emph{noncolinear}.
In other words, $\ket{\Phi}$ cannot be decomposed into another basis
\begin{equation}
\ket{\Phi}=\sum_{n}c_{n}\ket{s_{n}}\ket{a_{n}}\ket{e_{n}}\not\equiv\sum_{n}c_{n}^{\prime}\ket{s_{n}^{\prime}}\ket{a_{n}^{\prime}}\ket{e_{n}^{\prime}}
\end{equation}
The preferred basis of the apparatus is chosen by the interaction
with the environment, in that the projection $\widehat{P_{n}}=\ket{a_{n}}\bra{a_{n}}$
should commute with the interaction Hamiltonian $\left[H_{ae},\widehat{P_{n}}\right]=0$
, in order that any correlation of the measured system with the eigenstates
of a preferred apparatus observable $\widehat{O}=\sum_{n}\lambda_{n}\ket{a_{n}}\bra{a_{n}}$
is preserved.

The key point in solving the PBP is the use of the tridecompositional
uniqueness theorem. However, this does not necessarily require introducing
the environment. From the previous discussion we see that to conserve
total momentum, as long as a measurement type interaction happens,
there should always be another degree of freedom of the system $|p\rangle$
that carries the change due to the interaction, 
\begin{equation}
\ket{\Phi}=\sum c_{n}\ket{s_{n}}\ket{p_{0}}\ket{a_{0}}\stackrel{H_{sa}}{\longrightarrow}\sum c_{n}\ket{s_{n}}\ket{p_{n}}\ket{a_{n}}
\end{equation}
where $\ket{s_{n}}$, $\ket{p_{n}}$ are the states corresponding
to the two degrees of the system, and $\ket{a_{n}}$ is the state
of the macroscopic system. In realistic cases $|p\rangle$ is usually
used as the the apparatus, since the difference between $\ket{a_{0}}$
and $\ket{a_{n}}$ is negligible. The point here is that, as long
as the overlap is not exactly one, namely, $\{\ket{a_{n}}\}$ are
not collinear, the tridecompositional uniqueness theorem can be used.
This means that before the interaction between the apparatus and the
environment, the preferred basis has \emph{already} been determined
by the interaction between the system and apparatus. The preferred
basis is selected by the intrinsic mechanism of the measurement interaction,
or in other words, the internal configuration of the measurement apparatus,
instead of the environment. We name this solution as ``interaction
induced superselection'', (Inselection).

\section{Inseletion vs Einselection}

The advantage of Inselection over Einselection is significant. First
of all, Inselection is simpler, and more natural. When designing a
quantum measurement experiment, what we actually do is to find a mechanism
such that, if the input quantum state is $\ket{s_{1}}$, the output
state of the classical apparatus is $\ket{a_{1}}$, if input $\ket{s_{2}}$,
the output is $\ket{a_{2}}$, etc, and all these ${\ket{a_{n}}}$
can be differentiated without an additional quantum measurement. The
preferred bases in the PBP are just the states corresponding to the different
final configuration of the apparatus after coupling to different states
of the quantum system, as we expected. In order to abide by the third
postulate of QT, we choose this mechanism in which there will be a
third degree of freedom that carries the change due interaction between
the apparatus and the system. The three degrees of freedom become
entangled during this interaction so that the global state has only
one decomposition, and that guarantees we will end up with the preferred
basis of the apparatus. The preferred basis is already selected by
this measuring interaction, without the need to introduce the environment.
The role of the environment is to break the quantum coherence of the different states of the apparatus,
when they are coupled to the environment, rather than to select
the preferred basis. There is only one interaction in Inselection
after which the preferred basis is readily determined, while in Einselection,
which ignores the third degree of freedom, the preferred basis is
not determined until the second interaction with the environment. 

Here it will be beneficial to clarify a common misunderstanding of
decoherence and Einselection for the PBP, which are closely related. Decoherence
is a highly successful theory, which explains how a quantum system
loses its quantum coherence due to its interaction with the environment.
Decoherence deals with the the system-environment interaction. The
global system includes only the system and the environment and there
is no pre-defined apparatus. There is also a \emph{pointer basis} in
decoherence, but that is the basis of the environmental states entangled
with the system, which quickly become orthogonal and remains stable
under the interaction with other parts of the environment. On the
contrary, in the PBP Einselection serves to determine the preferred
basis of the \emph{apparatus}, by introducing a third party, the environment,
so that the tridecompositional uniqueness theorem can be used to rule
out other possible decompositions. In that case the preferred basis
is chosen by the interaction between the apparatus and the environment.

The introduction of the environment to select the preferred basis
is, not only redundant, but problematic. We should note that the PBP
\emph{only arises} when the expansion coefficient\emph{ }$c_{n}$ are
the same, otherwise the decomposition is unique. This means, according
to Einselection, the preferred basis is determined by the interaction
between the system and environment only for the special cases when
$c_{n}$ are the same, while for all other cases when the $c_{n}$
are distinct, the preferred basis is determined by another interaction,
the interaction between the apparatus and system. Thus the physical
mechanism determining the preferred basis is different only because
the superposition coefficients of the system are different. However,
in real world, there is hardly any difference between $c_{n}$ and
$c_{n}+\delta$, when $\delta$ is very small. In Inselection, on
the other hand, the preferred basis is determined by the interaction
between the system and apparatus for all cases, which is more consistent.

Moreover, Einselection implies that if we could control the interaction
between the apparatus and the environment, we can actually select
a different preferred basis than what naturally arises from the mechanism
of the measurement. This leads to serious problems. Let us consider
a common example in literature introducing the PBP\cite{schlosshauer_decoherence_2005}(this
example actually has a problem, see discussion later). Suppose we
have a spin in the state $\ket{z_{-}}_{1}=\frac{1}{\sqrt{2}}(\ket{x_{+}}_{1}-\ket{x_{-}}_{1})$.
Now this becomes entangled with another spin state, using the second
spin as the measurement apparatus. The composite system is

\begin{align*}
\ket{\psi}=\ket{z_{-}}_{1}\ket{x_{+}}_{2}= & \frac{1}{\sqrt{2}}(\ket{x_{+}}_{1}-\ket{x_{-}}_{1})\ket{x_{+}}_{2}\\
\stackrel{H_{sa}}{\longrightarrow} & \frac{1}{\sqrt{2}}(\ket{x_{+}}_{1}\ket{x_{-}}_{2}-\ket{x_{-}}_{1}\ket{x_{+}}_{2})\\
= & \frac{1}{\sqrt{2}}(\ket{z_{+}}_{1}\ket{z_{-}}_{2}-\ket{z_{-}}_{1}\ket{z_{+}}_{2})
\end{align*}

According to Einselection, if we can control the environment $\left[H_{ae},\pm\ket{z_{\pm}}_{2}\bra{z_{\pm}}_{2}\right]=0$,
so that \{$\ket{z_{\pm}}_{2})$\} is selected, after tracing out the
environment we will find the first spin with half probability of
being in $\ket{z_{+}}$ and half in $\ket{z_{-}}$. However, we actually
start in $\ket{z_{-}}_{1}$, thus this interaction no longer has the
function of measurement! Obviously, for Inselection such problem does
not exist.

Some readers may be confused about the apparatus in our Inselection
and the environment in Einselection, thinking that in Inselection
the apparatus is playing the role of the environment while the second
degree of freedom that carries the change is playing the role of apparatus
in Einselection. This is not really the case. The apparatus is a finite
object, and the interaction is finished after the system leaves the
apparatus. While the environment is the collection of all the other
degrees of freedom, such as the photons and air molecules scattering
around, and the interaction with the environment will last forever.
Moreover, in Einselection, the environment is usually factorizable,
which is not the case for the apparatus. 

Now let's return to the above example and analyze why it is problematic.
In a quantum measurement scheme, the apparatus should be macroscopic
so that the different states can be determined classically. However,
in the above problem the second spin is used as the measurement apparatus,
and to determine the state of this apparatus we need to make another
quantum measurement.This is not a valid quantum measurement. The correct
version is to replace the second spin with a macroscopic apparatus
such as a panel with a pointer. When a $\ket{x_{+}}$ spin passes
through the apparatus the pointer will point up, i.e.~it is described
by the state$\ket{a_{-}}$, and when a $\ket{x_{-}}$ spin passes
through it is described by $\ket{a_{+}}$ . If we sent a $\ket{z_{-}}$
spin through the apparatus, then the composite system evolves according
to 
\begin{align*}
\ket{\psi}=\ket{z_{-}}\ket{a_{+}}= & \frac{1}{\sqrt{2}}(\ket{x_{+}}-\ket{x_{-}})\ket{a_{+}}\\
\stackrel{H_{sa}}{\longrightarrow} & \frac{1}{\sqrt{2}}(\ket{x_{+}}\ket{a_{-}}-\ket{x_{-}}\ket{a_{+}})\\
= & \frac{1}{\sqrt{2}}(\ket{z_{+}}\ket{a_{\downarrow}}-\ket{z_{-}}\ket{a_{\uparrow}})
\end{align*}
with $\ket{a_{\downarrow}}:=\frac{1}{\sqrt{2}}(\ket{a_{+}}-\ket{a_{-}})$
and $\ket{a_{\uparrow}}:=\frac{1}{\sqrt{2}}(\ket{a_{+}}+\ket{a_{-}})$,
which corresponds to the macroscopic superposition of a pointer pointing
up and down. The PBP turns to be why in reality we always end
up with classical states $\ket{a_{+}}$ and $\ket{a_{-}}$, which
is in principle position dependent, instead of $\ket{a_{\downarrow}}$
and $\ket{a_{\uparrow}}$, a superposition of position bases. Our
answer is that this is determined by the mechanism of the measurement
interaction, which naturally yields localized states are we expected.
The Einselection answer to this is by resorting to classical mechanics,
that the interaction between the apparatus and environment depends
on position, so that $\left[H_{ae},\pm\ket{a_{\pm}}\bra{a_{\pm}}\right]=0$ (See 
e.g.~\cite{schlosshauer_decoherence_2005,zurek_environment-induced_1982}).
At this point the two approaches converge, since both $H_{sa}$ and
$H_{ae}$are fundamentally position dependent, but in Einselection
there is an unnecessary detour due to the introduction of environment.

\section{Conclusion}

We have shown that in a quantum measurement, the interaction between
the apparatus and the quantum system will entangle three degrees of
freedom, one system degree of freedom that we are interested in and
it is preserved by the interaction, one system degree of freedom that
carries the change due to the interaction, and the apparatus degree
of freedom which is always ignored. Considering all three degrees
of freedom the composite state only has one decomposition, and this
guarantees the apparatus will end up in the preferred basis which
is what we expect in daily experience. We also point out some problems
with the previous Einselection solution to the PBP, and this clarifies
a common misunderstanding of decoherence and the preferred basis problem.

\section{Acknowledgements}
The authors would like to thank  N. Babcock, C. Brukner, S.Das, P. Kwiat, J. Riedel, N. Sangouard, C.Simon, D.M. Tong and W. Wootters for helpful discussion.

\end{document}